\documentclass[12pt]{article}
\usepackage{amsmath}
\usepackage{graphicx}
\newcommand{\ave}[1]{\left\langle #1 \right\rangle }

\begin{document}

\title{\textbf{{Remarks on possible local parity violation in heavy ion collisions} }}
\author{Adam Bzdak$^{a,b}$, Volker Koch$^{a}$, and Jinfeng Liao$^{a}$ 
\\\\
$^{a}$ Lawrence Berkeley National Laboratory, 1 Cyclotron Road\\
MS70R0319, Berkeley, CA 94720, USA\thanks{e-mails: ABzdak@lbl.gov, VKoch@lbl.gov, JLiao@lbl.gov} \\
$^{b}$ Institute of Nuclear Physics, Polish Academy of Sciences\\
Radzikowskiego 152, 31-342 Krakow, Poland }
\maketitle

\begin{abstract}
In this note we discuss some observations concerning the possible local parity violation in heavy ion collisions recently announced by the STAR Collaboration. Our results can be
summarized as follows (i) the measured correlations for same charge pairs are mainly in-plain and not out of plane, (ii) if there is a parity violating component it is large
and, surprisingly, of the same magnitude as the background, and (iii) the observed dependence
of the signal on the transverse momentum ($p_t$) is consistent with a soft boost in $p_t$ and thus in line with expectations from the proposed chiral magnetic effect.

\vskip 0.6cm

\end{abstract}

\newpage

\section{Introduction}

Recently the STAR collaboration announced \cite{star-p} the results on
possible local parity violation in heavy ion collisions. In Ref. \cite{DK,DK-latest} it was argued
that in the hot dense matter created in heavy ion collisions local, instanton or sphaleron,
transitions to QCD vacua with different topological charge may result in metastable domains,
where parity is locally violated. 

In this paper we will solely concentrate on an analysis of the experimental results. We
will neither  attempt to provide alternative explanations for the observed correlations, such as
e.g. given in Ref. \cite{FW} nor will we discuss the likelihood that the proposed effect may occur
in a heavy ion collision. For detailed discussion of the underlying mechanism and the latest 
theoretical review of this problem we refer the reader to Ref. \cite{DK-latest}.
 
The phenomenon due to local parity violation, which is of relevance for the discussion here, is the so called chiral magnetic effect \cite{DK,DK-latest}. It leads to the separation of negatively
and positively charged particles along the system's angular momentum (or equivalently
the direction of the magnetic field) into two hemispheres separated by the
reaction plane. As a result, the system exhibits a electric current along the direction
of the angular momentum, and thus breaks parity locally in a given event. However, since instanton
(sphaleron) and anti-instanton (anti-sphaleron) transitions occur equally likely, the chiral
magnetic current is either aligned or anti-aligned with the angular momentum. As a result, the
expectation value of any parity odd observable, such as $\ave{\vec{j}_{CM} \vec{I}}$ vanishes. Here
$\vec{j}_{CM}$ is the chiral-magnetic current and $\vec{I}$ is the angular momentum.
Consequently, a direct measurement of parity violation even in a small subsystem is impossible.
However, one may attempt to identify the existence of these parity violation domains by studying the
fluctuations or the variance of a parity-odd observable. Since the variance of a parity-odd
observable is parity even, in principle other, genuinely parity-even, effects may contribute, and
one needs to separate those carefully before being able to draw any conclusions about the existence
of local, parity violating domains. 

\begin{figure}[h]
\begin{center}
\includegraphics[scale=1]{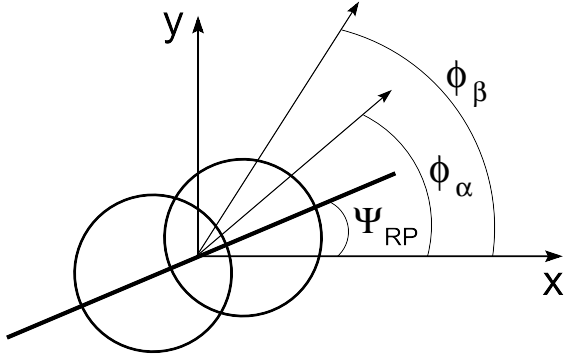}
\end{center}
\caption{The transverse plain in a collision of two heavy ions. $\Psi _{RP},$
$\protect\phi _{\protect\alpha }$ and $\protect\phi _{\protect\beta }$
denote the azimuthal angles of the reaction plane and produced charged
particles, respectively.}
\label{fig_rplain}
\end{figure}

In Ref. \cite{SV} Voloshin proposed a method to measure the variance of a parity odd observables. He
suggested to measure the following correlator $\left\langle \cos (\phi
_{\alpha }+\phi _{\beta }-2\Psi _{RP})\right\rangle $, where $\Psi _{RP},$ $%
\phi _{\alpha }$ and $\phi _{\beta }$ denote the azimuthal angles of the
reaction plane and produced charged particles respectively, see Fig. \ref{fig_rplain}.

As we will discuss in more detail in Section \ref{sec:integrated} this rather
involved correlation function has the advantage that correlations which are independent of the
reaction plane do not contribute. As a result, a large fraction of the expected background should
cancel. Recently the STAR collaboration has reported the measurement of the above correlation
function \cite{star-p}, both integrated over the entire acceptance as well as differential in
transverse momentum and pseudo-rapidity.

This paper is organized as follows. In the following section we will analyze the integrated STAR
result and will suggest additional measurements necessary to further clarify the situation. In the subsequent section, we will concentrate on the $p_t$ differential results and explore to which extend they are consistent with the expected soft phenomena due to the chiral magnetic effect.  

\section{The integrated signal}
\label{sec:integrated}

In Ref. \cite{star-p} the details of the STAR measurement are given. Among
other things STAR shows the results for $\left\langle \cos (\phi _{\alpha
}-\phi _{\beta })\right\rangle $ and for $\left\langle \cos (\phi _{\alpha
}+\phi _{\beta }-2\phi _{c})\right\rangle $, where $\phi _{\alpha ,\beta ,c}$
are the azimuthal angles of the produced charged particles. The paper gives
reasonable arguments that%
\begin{equation}
\left\langle \cos (\phi _{\alpha }+\phi _{\beta }-2\phi _{c})\right\rangle
=\left\langle \cos (\phi _{\alpha }+\phi _{\beta }-2\Psi _{RP})\right\rangle
v_{2,c},
\label{react_plane}
\end{equation}%
where $\Psi _{RP}$ is the angle of the reaction plane, and $v_{2,c}$
characterizes the elliptic anisotropy for the particle with angle $\phi _{c}$.

For the rest of the discussion we will assume that the  relation (\ref{react_plane}) is correct.
As a consequence, we will work in a frame where the reaction
plane is defined by the $x-z$ coordinates and where the $y$ direction is
perpendicular to the reaction plane. In other word we work in a frame where $%
\Psi _{RP}=0$, see Fig. \ref{fig_rplain}. Furthermore, since $\left\langle \cos (\phi
_{\alpha }-\phi _{\beta })\right\rangle $ is independent of the direction
of the reaction plane\footnote{%
Indeed $\left\langle \cos (\phi _{\alpha }-\phi _{\beta })\right\rangle
\equiv \left\langle \cos ([\phi _{\alpha }-\Psi _{RP}]-[\phi _{\beta }-\Psi
_{RP}])\right\rangle $}, it will be the same also in the frame where the
reaction plane is specified e.g., $\Psi _{RP}=0$. Thus within our frame we
have to consider the following two-particle correlations:%
\begin{eqnarray}
\left\langle \cos (\phi _{\alpha }-\phi _{\beta })\right\rangle
&=&\left\langle \cos (\phi _{\alpha })\cos (\phi _{\beta })\right\rangle
+\left\langle \sin (\phi _{\alpha })\sin (\phi _{\beta })\right\rangle , 
\notag \\
\left\langle \cos (\phi _{\alpha }+\phi _{\beta })\right\rangle
&=&\left\langle \cos (\phi _{\alpha })\cos (\phi _{\beta })\right\rangle
-\left\langle \sin (\phi _{\alpha })\sin (\phi _{\beta })\right\rangle .
\label{math}
\end{eqnarray}

STAR has measured both these correlation functions for same sign, $\left(
+,+\right) ,\left( -,-\right) $, and opposite sign, $\left( +,-\right) $,
pairs of charged particles. Qualitatively the data for $Au+Au$ collisions can be characterized as
follows.

\begin{itemize}
\item For same sign pairs:%
\begin{equation}
\left\langle \cos (\phi _{\alpha }+\phi _{\beta })\right\rangle
_{same}\simeq \left\langle \cos (\phi _{\alpha }-\phi _{\beta
})\right\rangle _{same}<0.
\end{equation}%
Using Eq. (\ref{math}) this implies%
\begin{eqnarray}
\left\langle \sin (\phi _{\alpha })\sin (\phi _{\beta })\right\rangle
_{same} &\simeq &0,  \notag \\
\left\langle \cos (\phi _{\alpha })\cos (\phi _{\beta })\right\rangle
_{same} &<&0.  \label{ss0}
\end{eqnarray}

\item For opposite sign pairs we find that%
\begin{eqnarray}
\left\langle \cos (\phi _{\alpha }+\phi _{\beta })\right\rangle _{opposite}
&\simeq &0  \notag \\
\left\langle \cos (\phi _{\alpha }-\phi _{\beta })\right\rangle _{opposite}
&>&0.
\end{eqnarray}

Again, using Eq. (\ref{math}), this means%
\begin{equation}
\left\langle \sin (\phi _{\alpha })\sin (\phi _{\beta })\right\rangle
_{opposite}\simeq \left\langle \cos (\phi _{\alpha })\cos (\phi _{\beta
})\right\rangle _{opposite}>0.
\end{equation}

\end{itemize}

\begin{figure}[h!!]
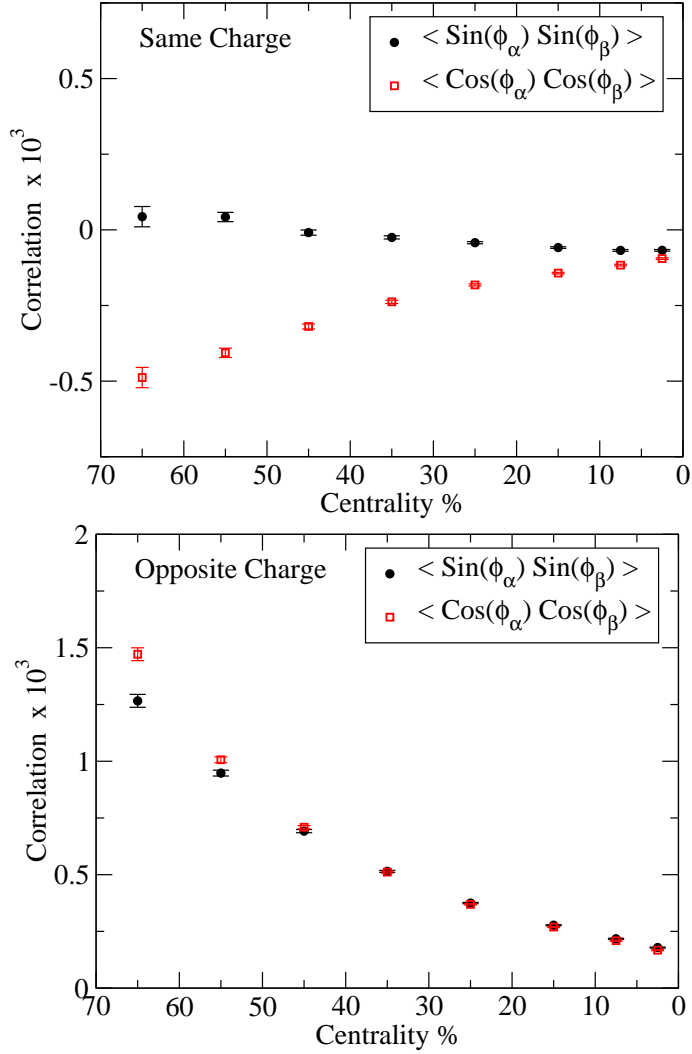

\begin{center}
\includegraphics[scale=0.5]{same.eps}
\includegraphics[scale=0.5]{opposite.eps}
\end{center}
\caption{Correlations in-plane $\left\langle \cos (\protect\phi _{\protect%
\alpha })\cos (\protect\phi _{\protect\beta })\right\rangle $ and out of
plane $\left\langle \sin (\protect\phi _{\protect\alpha })\sin (\protect\phi %
_{\protect\beta })\right\rangle $ for same and opposite charge pairs in $Au+Au$ collisions. As can be seen the correlations for same charge pairs are mainly in-plane.}
\label{fig_sscc}
\end{figure}

The actual data decomposed into the above components are shown in Fig. \ref{fig_sscc}.

The fact that for same charge pairs the sinus-term in Eq. (\ref{ss0}) (see Fig. \ref{fig_sscc}) is
essentially zero whereas the cosine term is finite, 
tells us that the observed correlations are actually \emph{in plane} rather than out of plane. This
is contrary to the expectation from the chiral magnetic effect, which results in same charge
correlation out of plane. In addition, since the cosine term is negative, the in-plane correlations
are stronger for back-to-back pairs than for small angle pairs. 
Second, we see that for opposite charge pairs the in- and out-of-plane correlations are virtually
identical. This is hard to comprehend, given that there is a sizable elliptic flow in these
collisions. At present, there is no simple explanation for neither of these observations.
However, they may be explained by a cluster model, which requires several, not unreasonable,
assumptions \cite{FW}. 

One may ask if there is room for a parity violating component if for the
same sign $\left\langle \sin (\phi _{\alpha })\sin (\phi _{\beta
})\right\rangle _{same}\simeq 0$, i.e. the signal is in-plane rather than out of plane. Following
the argument of Ref. %
\cite{star-p,SV}, we can always write%
\begin{equation}
\left\langle \sin (\phi _{\alpha })\sin (\phi _{\beta })\right\rangle
_{same}=B_{out}+P,  \label{P}
\end{equation}%
where $P$ is the part of the correlation which is caused be the parity
violation (at this stage we do not claim that $P\neq 0$) and $B_{out}$
represents all other contributions by correlations projected on the
direction perpendicular to the reaction plane. Denoting the correlations
in-plane $\left\langle \cos (\phi _{\alpha })\cos (\phi _{\beta
})\right\rangle _{same}$ by $B_{in}$ we obtain:%
\begin{eqnarray}
\left\langle \cos (\phi _{\alpha }+\phi _{\beta })\right\rangle _{same}
&=&[B_{in}-B_{out}]-P,  \notag \\
\left\langle \cos (\phi _{\alpha }-\phi _{\beta })\right\rangle _{same}
&=&[B_{in}+B_{out}]+P.
\end{eqnarray}

The advantage of $\left\langle \cos (\phi _{\alpha }+\phi _{\beta
})\right\rangle $ is obvious. The background is $B_{in}-B_{out}$, meaning
that all correlations that do not depend on the reaction plane
orientation cancel. The STAR collaboration studied many known sources of reaction
plane dependent correlations and all effects produce $B_{in}-B_{out}$ which is much
smaller than the observed signal. We note, however, that at present the background
is not understood since none
of the present models is able to explain the value of $\left\langle \cos
(\phi _{\alpha }-\phi _{\beta })\right\rangle $.
 
Following the above argument, however, immediately implies that (using Eq. (\ref{ss0}) and Eq.
(\ref{P}))
\begin{equation}
P\simeq -B_{out}\simeq -B_{in}, 
\label{end}
\end{equation}%
i.e., the parity violating effect has to be precisely of the same
magnitude as all other, standard correlations. This
relation is quite an unexpected coincidence. It means that the parity signal
is quite strong, and consequently should also be  visible in $\left\langle
\cos (\phi _{\alpha }-\phi _{\beta })\right\rangle _{same}$ if the
background is well understood.

In our view, it is mandatory to explore if the relation, Eq. (\ref{end}), is just a 
coincidence or an indication of potential problems with the present
interpretation of the data. To answer this question it is essential to
analyze the correlation function $\left\langle \cos (\phi _{\alpha
}-\phi _{\beta })\right\rangle _{same}$  differentially in transverse momentum and
pseudo-rapidity as it has been  already done for  $\left\langle \cos (\phi _{\alpha }+\phi
_{\beta })\right\rangle _{same}$. Should relation (\ref{end}) persist also for the differential
correlations, one would have to conclude that the proposed parity violating effect is not seen in the data.

\section{Transverse momentum dependence}

The STAR collaboration has also presented \cite{star-p} the measurement of $\left\langle \cos
(\phi _{\alpha }+\phi _{\beta })\right\rangle $ in mid-central $Au+Au$ collisions as a
function of $p_{+}=(p_{t,\alpha }+p_{t,\beta })/2$ and $p_{-}=\left|
p_{t,\alpha }-p_{t,\beta }\right| $, where $p_{t,\alpha }$ and $p_{t,\beta }$
are the absolute values of the particles momenta. Qualitatively the data can
be characterized as follows.\footnote{In the present Section we are only interested in the $p_{t}$ dependence of the signal, not in the overall normalization.}

\begin{itemize}
\item For same sign pairs in the range $0<p_{+},p_{-}<2.2$ GeV:%
\begin{eqnarray}
\left\langle \cos (\phi _{\alpha }+\phi _{\beta })\right\rangle _{p_{+},%
\text{ }same} &\propto &p_{+}  \notag \\
\left\langle \cos (\phi _{\alpha }+\phi _{\beta })\right\rangle _{p_{-},%
\text{ }same} &\simeq &const.  \label{data-pt}
\end{eqnarray}

\item For opposite sign pairs  the signal vs $p_{+}$
and $p_{-}$ is consistent with zero.
\end{itemize}

One would  expect \cite{star-p,DK} that the parity violating signal should be a soft, low
$p_{t}$ phenomenon. Thus the observed increase of the signal for same sign
pairs with $p_{+}$ seems to be inconsistent with the chiral magnetic effect. As we will
show such a conclusion is not necessarily correct and the true signal may indeed be
consistent with the expected low $p_t$ dynamics.

Indeed, by definition%
\begin{equation}
\left\langle \cos (\phi _{\alpha }+\phi _{\beta })\right\rangle =\frac{%
N_{corr}}{N_{all}},
\end{equation}%
where $N_{corr}$ is the number of correlated pairs [via $\cos (\phi _{\alpha
}+\phi _{\beta })$] and $N_{all}$ is the number of all pairs. The latter can
be easily approximated by [$p_{+}=(p_{t,\alpha }+p_{t,\beta })/2$ and $%
p_{-}=\left| p_{t,\alpha }-p_{t,\beta }\right| $]:%
\begin{eqnarray}
N_{all}\left( p_{+}\right) &\propto &\int d^{2}p_{t,\alpha }d^{2}p_{t,\beta
}\exp \left( -\frac{p_{t,\alpha }}{T}\right) \exp \left( -\frac{p_{t,\beta }%
}{T}\right) \delta \left( 2p_{+}-[p_{t,\alpha }+p_{t,\beta }]\right)  \notag
\\
&\propto &p_{+}^{3}e^{-2p_{+}/T}  \label{all-sum}
\end{eqnarray}%
and%
\begin{eqnarray}
N_{all}\left( p_{-}\right) &\propto &\int d^{2}p_{t,\alpha }d^{2}p_{t,\beta
}\exp \left( -\frac{p_{t,\alpha }}{T}\right) \exp \left( -\frac{p_{t,\beta }%
}{T}\right) \delta \left( p_{-}-\left| p_{t,\alpha }-p_{t,\beta }\right|
\right)  \notag \\
&\propto &T^{2}e^{-(p_{-}/T)}\left( p_{-}+T\right) ,  \label{all-dif}
\end{eqnarray}%
where in the following calculations we take $T=0.22$ GeV.\footnote{%
It corresponds to the average transverse momentum of the pions $\left\langle
p_{t}\right\rangle =0.45$ GeV.}

The calculated distributions of all pairs vs $(p_{t,\alpha }+p_{t,\beta })/2$
and $\left| p_{t,\alpha }-p_{t,\beta }\right| $ are presented in Fig. \ref%
{fig_pairs}. It is worth noticing that both functions are concentrated in
the small $p_{t}$ region, reflecting typical thermal distributions for $p_{-}$ and $p_{+}$.
Due to the soft nature of chiral magnetic effect, one expects that the
distributions in $p_{-}$ and $p_{+}$ for the correlated particles should not differ much from the
underlying thermal distributions. This is indeed the case as we will demonstrate next.
\begin{figure}[h!!]
\begin{center}
\includegraphics[scale=0.9]{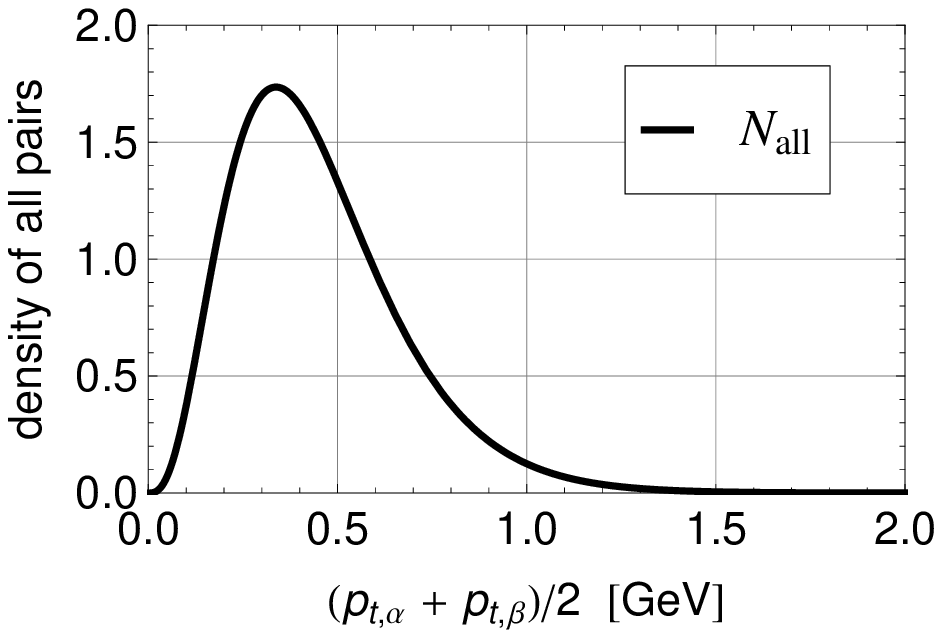} %
\includegraphics[scale=0.9]{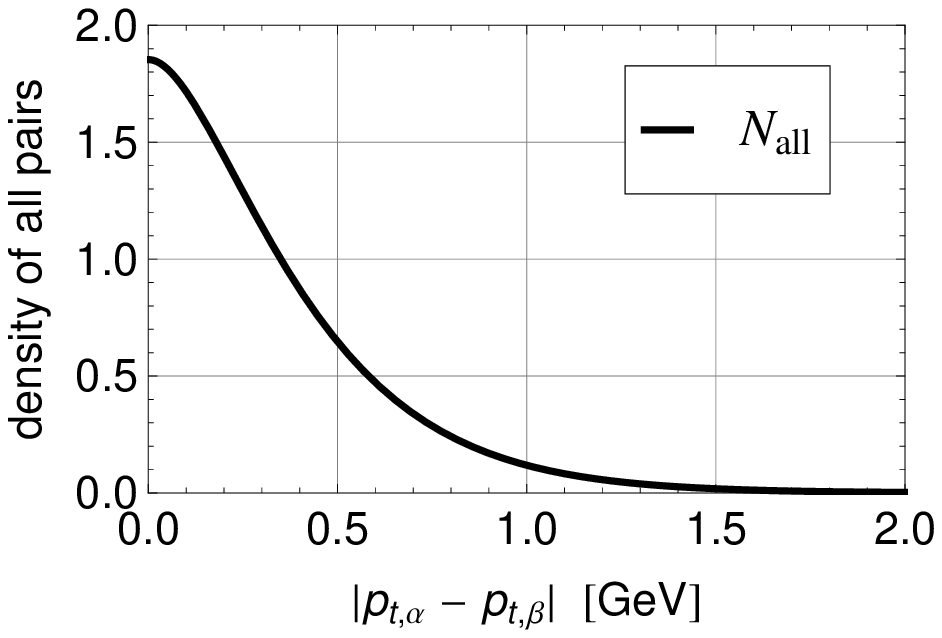}
\end{center}
\caption{The distributions of all charge pairs vs. $(p_{t,\protect\alpha %
}+p_{t,\protect\beta })/2$ and $\left| p_{t,\protect\alpha }-p_{t,\protect%
\beta }\right| $, respectively.}
\label{fig_pairs}
\end{figure}

In order to estimate the distribution of correlated same sign
pairs it is sufficient to multiply Eq. (\ref{data-pt}) by the expressions (\ref%
{all-sum}) and (\ref{all-dif}), respectively. Consequently we obtain 
\begin{eqnarray}
N_{corr}(p_{-}) &\propto &N_{all}\left( p_{-}\right),  \notag \\
N_{corr}(p_{+}) &\propto &p_{+}N_{all}(p_{+}) .
\end{eqnarray}

\begin{figure}[h!!]
\begin{center}
\includegraphics[scale=1.0]{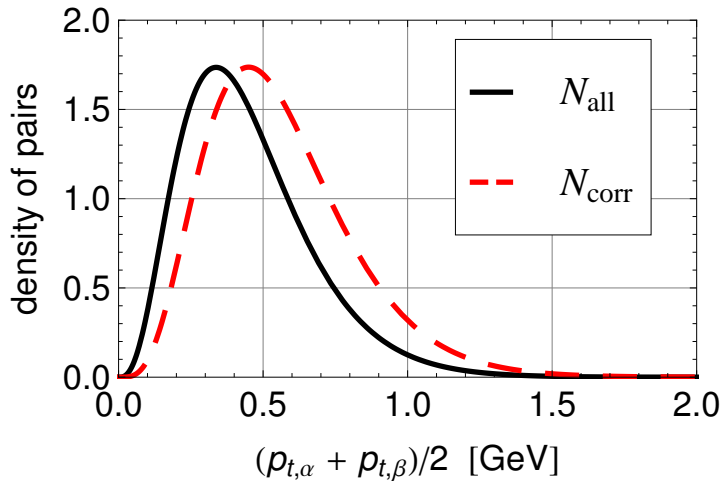}
\end{center}
\caption{Distribution of all pairs (solid line) compared with the distribution of same sign
correlated pairs (dashed line). Both functions are concentrated in the low $p_{t}$ region. 
}
\label{fig_boost}
\end{figure}

As can be seen the dependence of the number of correlated same pairs vs 
$\left| p_{t,\alpha }-p_{t,\beta }\right| $ is identical to the dependence
of all pairs presented in Fig. \ref{fig_pairs}. Clearly the signal is
concentrated in the low $p_{t}$ region and indeed is unchanged from a thermal distribution. 
In Fig. \ref{fig_boost} the
dependence of the number of same sign pairs vs $(p_{t,\alpha }+p_{t,\beta
})/2$ is compared with the dependence of all pairs (previously shown in Fig. %
\ref{fig_pairs}). We find that the momenta of correlated particles are
slightly shifted to the higher $p_{t}$ and the shape is roughly similar. The momentum shift required
by the data is $\delta p_{+} \simeq 150 \,\rm MeV$ which could conceivably be due to the large
magnetic field, although it is somewhat on the high end of what one would naively expect from
electromagnetic phenomena. 

\section{Conclusions}

In this note we have discussed several aspects of the recent measurement of possible local parity violation in $Au+Au$ collisions by the STAR Collaboration. We made the following three observations:

(i) For particles with the same charge STAR sees large negative correlations in-plain $\left\langle \cos(\phi _{\alpha })\cos (\phi _{\beta })\right\rangle _{same}$ and very small
correlations out of plain 
$\left\langle \sin (\phi _{\alpha })\sin (\phi_{\beta })\right\rangle _{same}$. For opposite sign correlations in-plane
and out-plain are both positive and of the same magnitude.

(ii) If there is indeed a parity violating component in the STAR data  it has to be of the
same magnitude as all other, ``trivial' correlations projected on the
direction perpendicular to the reaction plane. This may be a pure coincidence or an
indication that the present interpretation of the data as a signal for local parity violation needs
to be revised. To investigate this problem in more detail we
need differential distribution (vs pseudo-rapidity or transverse momenta) of $%
\left\langle \cos (\phi _{\alpha }+\phi _{\beta })\right\rangle $ and $%
\left\langle \cos (\phi _{\alpha }-\phi _{\beta })\right\rangle $ at the
same time.

(iii) We have also argued that the distribution of the number of correlated pairs
is concentrated in the low $p_{t}$ region i.e. $p_{t}<1$ GeV. It is not
inconsistent with the predictions of the chiral magnetic effect.

At present, the data from the STAR Collaboration does not allow for a definitive conclusion about
the presence of local parity violation. The measurement of the correlation function $
\left\langle \cos (\phi _{\alpha }-\phi _{\beta })\right\rangle $ differential in transverse
momentum and pseudo-rapidity is absolutely essential to further distinguish between trivial
correlations and those due to the chiral magnetic effect. 

\bigskip

\textbf{Acknowledgments}

We thank D. Kharzeev, L. McLerran, S. Voloshin, and F. Wang for useful discussions. 
This work was supported in part by the Director, Office of Energy Research, Office of High Energy and Nuclear Physics, Divisions of Nuclear Physics, of the U.S. Department of
Energy under Contract No. DE-AC02-05CH11231 and by the Polish Ministry of
Science and Higher Education, grant No. N202 125437, N202 034 32/0918. A.B.
also acknowledges support from the Foundation for Polish Science (KOLUMB
program).


\begin{thebibliography}{99}

\bibitem{star-p} STAR Collaboration, B.I. Abelev et al., e-Print:
arXiv:0909.1739 [nucl-ex]; arXiv:0909.1717 [nucl-ex].

\bibitem{DK} K. Fukushima, D.E. Kharzeev and H.J. Warringa, Phys. Rev. D78 (2008) 074033; D.E.
Kharzeev, L.D. McLerran and H.J. Warringa, Nucl. Phys. A803 (2008) 227.

\bibitem{DK-latest} D.E. Kharzeev, e-Print: arXiv:0911.3715 [hep-ph].

\bibitem{FW} F. Wang, e-Print: arXiv:0911.1482 [nucl-ex].

\bibitem{SV} S.A. Voloshin, Phys. Rev. C70 (2004) 057901.
\end{thebibliography}
\end{document}